\documentclass[aps,prb,reprint,superscriptaddress,notitlepage]{revtex4-2}
\usepackage{times,amsmath,amsfonts,amssymb,mathrsfs,graphics,graphicx,color,comment,bm}
\usepackage[next]{inputenc}
\usepackage[dvips]{epsfig}
\usepackage{hyperref}
\usepackage{pdfsync}
\usepackage{appendix}
\usepackage{textcomp}
\usepackage{lipsum}  
\usepackage{gensymb}
\usepackage{graphicx}
\usepackage{upgreek}

\newcommand{\antisky}{Mn$_{1.4}$Pt$_{0.9}$Pd$_{0.1}$Sn}
\begin{document}

\title{Spin dynamics in bulk MnNiGa and \antisky~investigated by muon spin relaxation}

\author{M. N. Wilson}
\address{Durham University, Department of Physics, South Road, Durham, DH1 3LE, United Kingdom}
\author{T. J. Hicken}
\address{Durham University, Department of Physics, South Road, Durham, DH1 3LE, United Kingdom}
\author{M. Gomil\v{s}ek}
\address{Jo\v{z}ef Stefan Institute, Jamova c. 39, SI-1000 Ljubljana, Slovenia}
\address{Durham University, Department of Physics, South Road, Durham, DH1 3LE, United Kingdom}
\author{A. \v{S}tefan\v{c}i\v{c}}
\address{Department of Physics, University of Warwick, Coventry CV4 7AL, United Kingdom}
\author{G. Balakrishnan}
\address{Department of Physics, University of Warwick, Coventry CV4 7AL, United Kingdom}
\author{J. C. Loudon}
\address{Department of Materials Science and Metallurgy, University of Cambridge, 27 Charles Babbage Road, Cambridge, CB3 0FS UK}
\author{A. C. Twitchett-Harrison}
\address{Department of Materials Science and Metallurgy, University of Cambridge, 27 Charles Babbage Road, Cambridge, CB3 0FS UK}
\author{F. L. Pratt}
\address{ISIS Pulsed Neutron and Muon Facility, STFC Rutherford Appleton Laboratory, Harwell Oxford, Didcot OX11 OQX, United Kingdom}
\author{M. Telling}
\address{ISIS Pulsed Neutron and Muon Facility, STFC Rutherford Appleton Laboratory, Harwell Oxford, Didcot OX11 OQX, United Kingdom}
\author{T. Lancaster}
\address{Durham University, Department of Physics, South Road, Durham, DH1 3LE, United Kingdom}

\begin{abstract}

  We report the results of muon-spin relaxation and magnetometry investigations of bulk \antisky~and MnNiGa, two materials that have been proposed to host topological magnetic states in thin lamellae (antiskyrmions for \antisky~and biskyrmions for MnNiGa), and show spin reorientation transitions in the  bulk. Our measurements reveal dynamic fluctuations surrounding the magnetic phase transitions in each material. In particular, we demonstrate that the behavior approaching the higher-temperature transitions reflects a  decrease in the frequency of dynamics with temperature.  At low temperatures the two systems both show spin dynamics over a broad range of frequencies that persist below the respective spin reorientation transitions. 

\end{abstract}

\maketitle

\section{Introduction}
Topological magnetic objects such as skyrmions are of fundamental interest in condensed matter physics \cite{Nagaosa2013,Lancaster2019}. These states arise from a complicated hierarchy of interactions and have unique topology that brings about significant energetic stability and interesting physical effects. Recently, the known variety of such topological excitations has been expanded by the  observation of two new states in thinned lamellae: biskyrmions \cite{Yu2014}, and antiskyrmions \cite{Nayak2017}. Biskyrmions are reported to consist of a bound state of two Bloch skyrmions with the same chirality, and hence have a topological charge of $N = 2$, while antiskyrmions are objects where the winding varies from N\'{e}el-type to Bloch-type around the circumference of the object, and have a topological charge of $N = -1$~~~\cite{Lancaster2019}.  

Biskyrmions have been reported in the layered manganite La$_{2-2x}$Sr$_{1+2x}$Mn$_2$O$_7$ \cite{Yu2014} and in certain compositions of hexagonal MnNiGa \cite{Wang2016,Peng2017}, while antiskyrmions have so far been reported in \antisky~\cite{Nayak2017} and related centrosymmetric Heusler systems \cite{Jena2020,Kumar2020}. Indirect evidence for the presence of a topologically nontrivial state has been reported for these materials, and other Heusler materials related to the antiskyrmion hosts \cite{Swekis2019, Rana2016, Liu2018, Li2018}, however direct evidence for both the biskyrmion and antiskyrmion states has thus far relied on Lorentz Transmission Electron Microscopy (LTEM).
However, this identification is controversial, as further LTEM and X-ray holography measurements of MnNiGa suggest that the LTEM images can be explained as reflecting a more conventional type-II magnetic bubble state \cite{Loudon2019},  while  theoretical studies continued to support the existence of biskyrmions in centrosymmetric magnetic films \cite{Capic2019}. Nevertheless, these materials remain of interest as a result of their complex magnetic behavior.

Both MnNiGa and \antisky~undergo two magnetic transitions with decreasing temperature \cite{Wang2016,Nayak2017}, with a ferromagnetic state occurring below $T_{\textrm{C}} = 400$~K (\antisky) or 350~K (MnNiGa), and a lower-temperature transition below 250~K in both materials that is accompanied by a change in magnetic moment. In \antisky, the transition is suggested to be a spin reorientation by analogy with similar materials such as Mn$_{1.4}$PtSn \cite{Nayak2017,Vir2019} and Mn$_2$RhSn \cite{Meshcheriakova2014}. Neutron diffraction measurements on MnNiGa show that the low-temperature transition in this material
 also involves spin reorientation, in this case introducing an antiferromagnetic component to create a canted non-collinear state \cite{Xu2019, Shiraishi1999}. 

Topological Hall effect anomalies have been reported in both materials. Below $T_{\textrm{C}}$ in 50~$\upmu$m-thick polycrystalline samples of MnNiGa, from 0.2~T $<\mu_0H <$ 1~T, a topological Hall signal is observed, consistent with the region in which the biskyrmion state is reported in sub-$\upmu$m thick lamellae \cite{Wang2016}. In contrast, in \antisky~a topological Hall signal is only reported below the spin-reorientation transition for $T < 150$~K \cite{Kumar2020} in bulk samples. This suggest that some sort of topologically non-trivial state exists in bulk samples of both materials, and not only in thin films or lamellae.

In this paper we present a study of the magnetic behavior of bulk MnNiGa and \antisky~using a combination of magnetometry and muon-spin relaxation ($\mu$SR). Our measurements reveal dynamic fluctuations around the two transitions in these materials that appear as a function of temperature. Further, the $\mu$SR underscores the difference in the nature of the fluctuating magnetism occurring between the lower spin reorientation transitions. The investigation of these fluctuations sheds light on the bulk behavior of these materials and helps put the observation of topologically non-trivial magnetic states in thin lamellae in context.

\section{Methods}
We performed muon-spin relaxation ($\mu$SR) and magnetometry measurements on samples of \antisky~and MnNiGa [nominally (Mn$_{0.5}$Ni$_{0.5}$)$_{65}$Ga$_{35}$] synthesized at the University of Warwick, UK. These samples consist of polycrystalline boules prepared by arc-melting the components together. The \antisky~boule is known to have regions of twinned crystallites that might introduce a preferred orientation to the sample, rather than having a fully random distribution of orientations. 

For the \antisky~$\mu$SR measurements, we used two halves of the boule with circular cross section, weighing a total of 1.217~g, with a total cross-sectional area of approximately 0.9~cm$^2$, while for the MnNiGa measurements we used a single boule of mass 0.907~g and circular cross-sectional area of 0.6~cm$^2$. $\mu$SR measurements were performed on the HiFi spectrometer at the STFC-ISIS pulsed muon source. The MnNiGa measurements were performed in two separate experimental runs, with the 0.01~T and 0.3~T measurements conducted first, and the other measurements later. Small differences in the setup of these experiments led to a difference in the observed high-temperature relaxing asymmetries.  For the magnetometry measurements, we cut a 8.7~mg piece from the \antisky~boule and a 4.8~mg piece from the MnNiGa boule, and measured them using a Quantum Design MPMS. AC magnetometry measurements were collected with a drive field of 0.3~mT at 111~Hz for \antisky~and 0.35~mT at 113~Hz for MnNiGa. 

\section{Results and Discussion}

\subsection{MnNiGa}

\subsubsection{Magnetometry}

\begin{figure}[h!]
\includegraphics[width=\columnwidth]{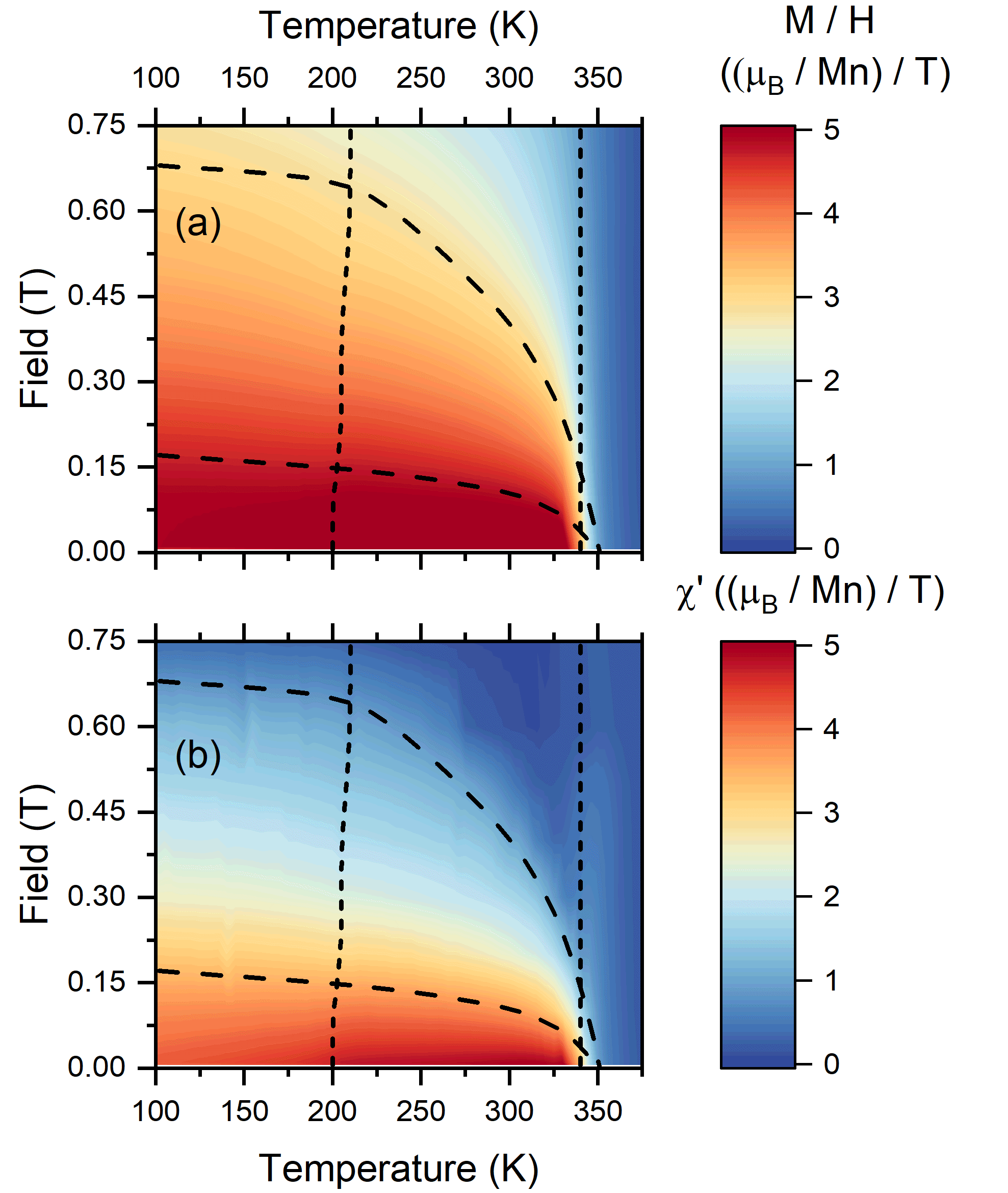}
\caption{Phase diagrams for MnNiGa measured through fixed magnetic field temperature sweeps on warming, after cooling in zero field. (a) DC magnetization, plotted as magnetization divided by applied field; (b) real part of the AC susceptibility ($\chi'$). The lines show transitions determined by DC magnetometry temperature scans (short dashed lines, maxima in d$M$/d$T$ for low-temperature transition, minima for the higher temperature transition), and AC magnetometry (long dashed lines, minima in d$\chi'$/d$H$).}
\label{fig:PD}
\end{figure}

\begin{figure*}[t]
\includegraphics[width=\textwidth]{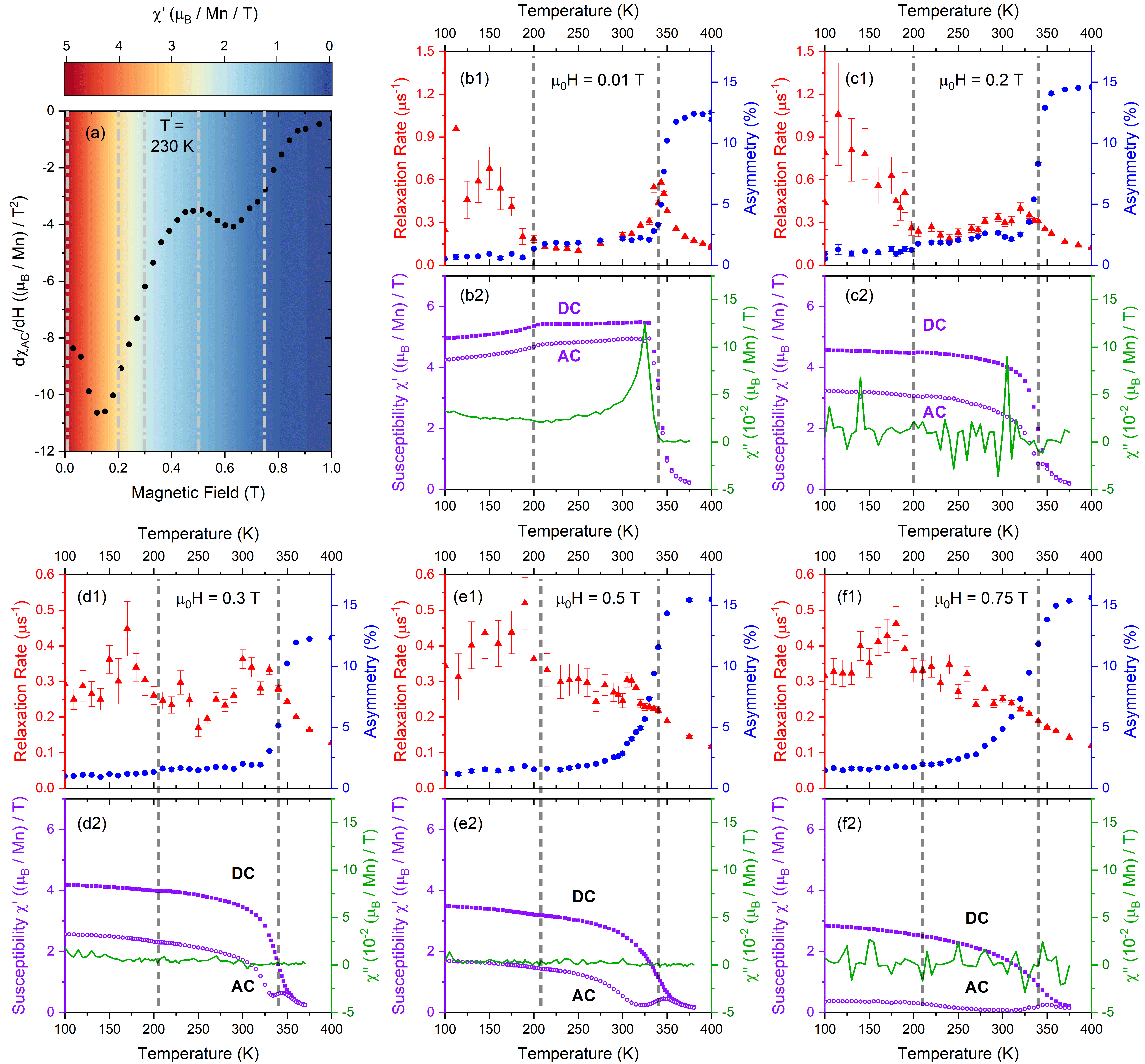}
\caption{Magnetometry and $\mu$SR data for MnNiGa measured at several magnetic fields as a function of temperature after zero-field cooling. (a) Derivative of the real part of the AC susceptibility $d\chi'/dH$ (black circles), as a function of field after ZFC to 230~K, and $\chi'$ as the colormap. (b)--(f), subpanel (1): measured $\mu$SR relaxation rate ($\lambda$, red triangles, left axis), the relaxing asymmetry ($A_\mathrm{r}$, blue circles, right axis); and  panel (2): DC magnetization plotted as $M/H$ (in purple squares, left axis), $\chi'$ (purple open circles, left axis), and the imaginary part of the AC susceptibility ($\chi''$, green solid line, right axis). Gray dashed lines in (b)--(f): temperatures of magnetic transitions determined from peaks in d$M$/d$T$ at each magnetic field (maxima for the low-temperature transition, minima for the higher temperature transition). The white dashed lines in (a) indicate the magnetic fields where panels (b)--(f) were measured.}
\label{fig:MnNiGa}
\end{figure*}

\begin{figure}[h!]
\includegraphics[width=0.8\columnwidth]{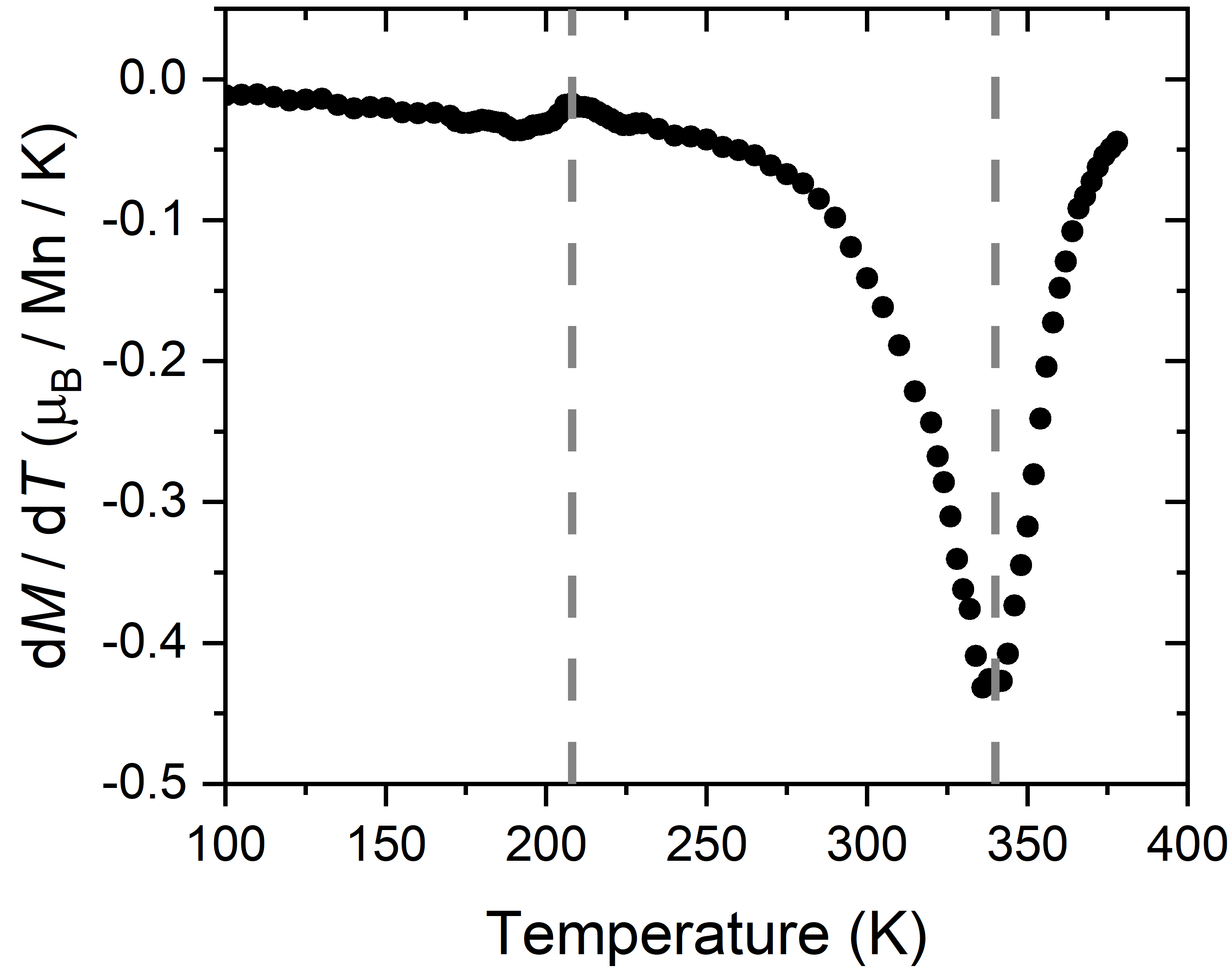}
\caption{d$M$/d$T$ for MnNiGa measured at 0.5~T. Grey dashed lines: transitions at 208 and 340~K determined from the extrema. }
\label{fig:dMdT}
\end{figure}

Figure~\ref{fig:PD} shows magnetic field--temperature phase diagrams for MnNiGa collected from (a) zero-field cooled (ZFC) temperature scans of DC magnetization (plotted as magnetization divided by field $M/H$),  and (b) the real part of the AC susceptibility ($\chi'$). (Examples of these measurements are shown in Fig.~\ref{fig:MnNiGa}.)
The DC magnetization shows a  transition at $T_{\textrm{C}} = 340$~K, characterized by a step-like increase in $M$ [also evident in e.g.~Fig.~\ref{fig:MnNiGa}(b2)], which remains nearly unchanged by applied magnetic field $H$. The AC magnetometry also shows a feature at this temperature that does not significantly change with $H$, showing up as a kink or weak maximum in $\chi'$ (most obviously at higher field values). This sort of feature has been seen in other materials such as the skyrmion host Cu$_2$OSeO$_3$, and interpreted as the transition from a field polarized state to a high temperature paramagnetic state \cite{Qian2016}.
There is also a second high-temperature feature in $\chi'$ that is not resolved in the $M/H$ data. This shows up as a sharp step-like increase in $\chi'$ that rapidly decreases in temperature as the field is increased, reminiscent of the transition from the conical to field polarized states in chiral helimagnets \cite{Qian2016}.
Below 200~K, the data show a small change in both magnetization and $\chi'$, suggesting a second magnetic transition 
occurring at all magnetic fields, consistent with previous reports. The transition temperature is determined from a peak in d$M$/d$T$ (see Fig.~\ref{fig:dMdT}).

Figure~\ref{fig:MnNiGa}(a) shows the magnetic field derivative of the real part of the AC susceptibility $d\chi'/dH$ at 230~K which has two pronounced minima at around 0.12~T and 0.62~T (data at other temperatures show similar behavior). This suggests that there is a crossover in magnetic behavior as a function of field, rather than a smooth evolution. The minimum at 0.62~T likely corresponds to the transition to a field polarized state, while the origin of the low-field transition is less clear. A similar series of two transitions was seen in MnNiGa in topological Hall effect measurements \cite{Wang2016}, and reported to arise from a low-field stripe/helical state, with biskyrmions reported at intermediate fields between the two transitions. Although, these states have only been reported in thin lamellae \cite{Wang2016}, our observation of a similar phase diagram in bulk samples suggests that these transitions may persist in the bulk.

\subsubsection{$\mu$SR}
To investigate the local magnetism in MnNiGa, we measured longitudinal field $\mu$SR. In this technique, spin polarized muons (with spins initially antiparallel to the longitudinal field $B_{\textrm{L}}$) are implanted into the material, where they then precess in the local magnetic field before decaying.
The average spin polarization of the muon ensemble is determined by measuring
the asymmetry function $A(t)=[F(t) - B(t)]/[F(t)+B(t)]$,
where $F$ represents the number of decay positrons measured forward of the initial muon spin direction and $B$ is the number measured in the backwards direction. 
In an ordered magnet there are typically two principal contributions to the time evolution of the asymmetry: (i) a  contribution arising from muons coherently precessing in the internal field directed perpendicular to the initial muon spin direction, and (ii)  a contribution arising from muon spins initially parallel to the local magnetic field dephased by time-dependent fluctuations. he first of these contributions would often result in damped oscillations in the asymmetry. However, the time resolution of a pulsed muon source limits our  ability to resolve these features, and this contribution gives rise to a fraction of `missing asymmetry' that relaxes too quickly to resolve. In contrast, the second contribution results in a slow decay of the asymmetry, determined by the spectral density $S(\omega)$
of local field fluctuations in the material, which is a Fourier transform of the local field-field autocorrelation function $\gamma_{\mu}^{2}\langle \delta B(\tau)\delta B(0)\rangle$, where $\gamma_{\mu} = 2\pi \times 135.5$~MHz/T is the muon
gyromagnetic ratio and $\delta B$ is the fluctuating field at the muon site.
The fluctuations will relax the muon spins most effectively when there is spectral density at the frequency associated with the applied longitudinal field
$\omega_0 = \gamma_{\mu} B_{\mathrm{L}}$ ($\approx 8$ - $600$~Mrad / s in this work, depending on the applied field $B_{\mathrm{L}}$), giving a longitudinal relaxation rate in the long-time limit of
\begin{equation}
  \lambda = \gamma_{\mu}^{2}\int_{0}^{\infty}\mathrm{d}\tau\cos\omega_{0}\tau\left[
\langle \delta B_{x} (\tau)\delta B_{x}(0)\rangle + \langle \delta B_{y}(\tau)\delta B_{y}(0)\rangle
  \right],
\end{equation}
where $\delta B_{x}$ and $\delta B_{y}$ represent field fluctuations perpendicular to the direction of the applied field $B_{\mathrm{L}}$ \cite{Yaounc2010}.

\begin{figure}
\includegraphics[width=\columnwidth]{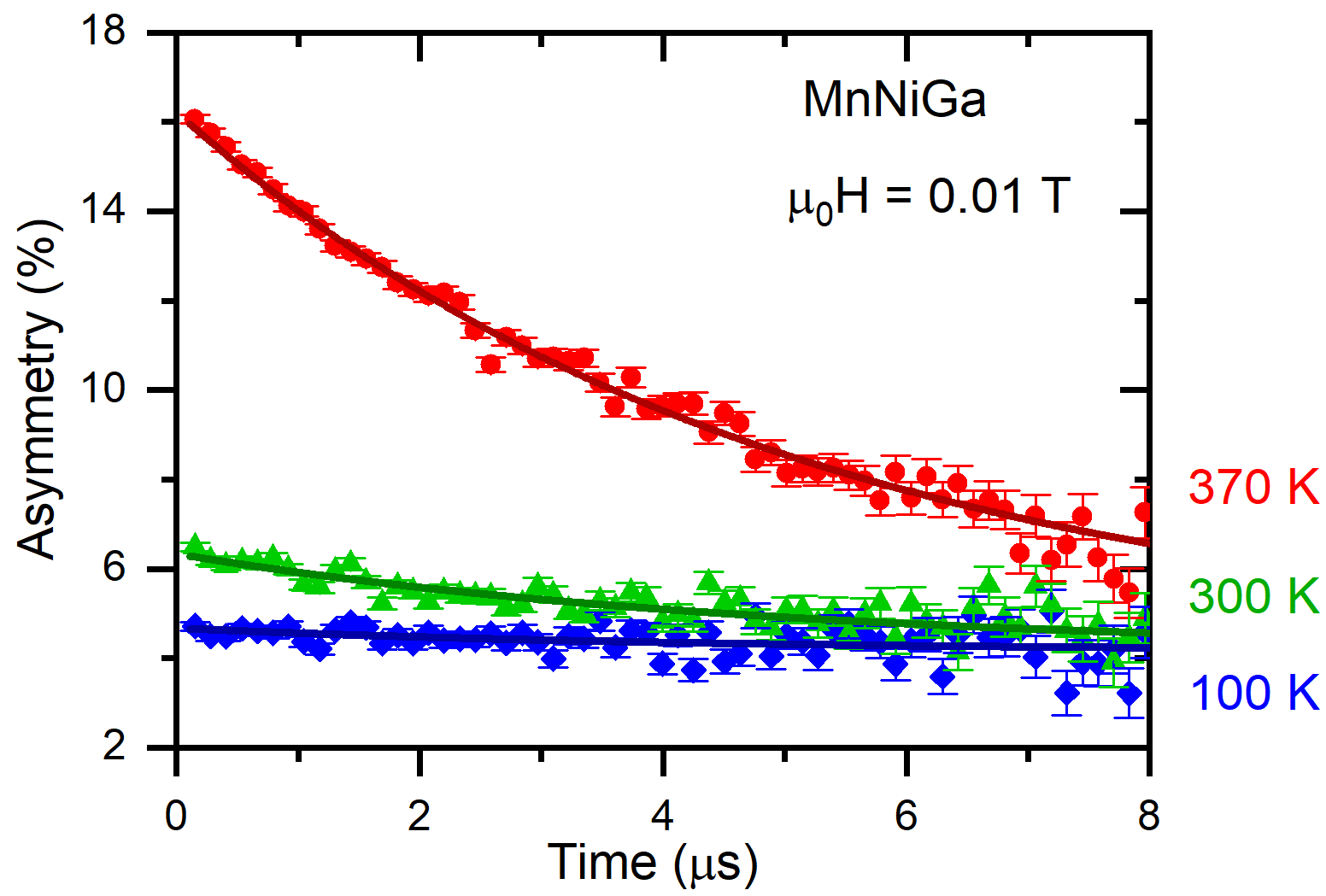}
\caption{Representative $\mu$SR data for MnNiGa measured  in a longitudinal field of 0.01~T at 100~K (blue), 300~K (green), and 370~K (red). Solid points show measured data; lines show fits to Eq.~\ref{eq:1}.}
\label{fig:data_MnNiGa}
\end{figure}

Figure~\ref{fig:data_MnNiGa} shows representative $\mu$SR data for the MnNiGa sample, measured in a longitudinal field of 0.01~T. The behavior is qualitatively similar at other measured fields. At each temperature the data can be well described by a single exponentially relaxing component, which relaxes to a temperature-independent baseline (4.2\% for the field shown). This behavior is characteristic of muons dephased  by dynamic fluctuation of magnetic moments in the material, and therefore provides information about how the dynamics vary across the phase diagram. We fit these data to
\begin{equation}
A(t) = A_\mathrm{b} + A_\mathrm{r} \mathrm{exp}(-\lambda t),
\label{eq:1}
\end{equation}
where $A_\mathrm{b}$ is the baseline asymmetry, $A_\mathrm{r}$ is the relaxing asymmetry, and $\lambda$ is the relaxation rate. The results of this fitting are shown in Fig.~\ref{fig:MnNiGa}, along with the corresponding temperature scans from magnetometry at equivalent magnetic fields. Dashed lines in Fig.~\ref{fig:MnNiGa}(a) show the magnetic fields where we measured $\mu$SR spectra, showing that we have one measurement in the low-field state, three spanning the intermediate field region, and one in the field-polarized state.

\begin{figure}
\includegraphics[width=\columnwidth]{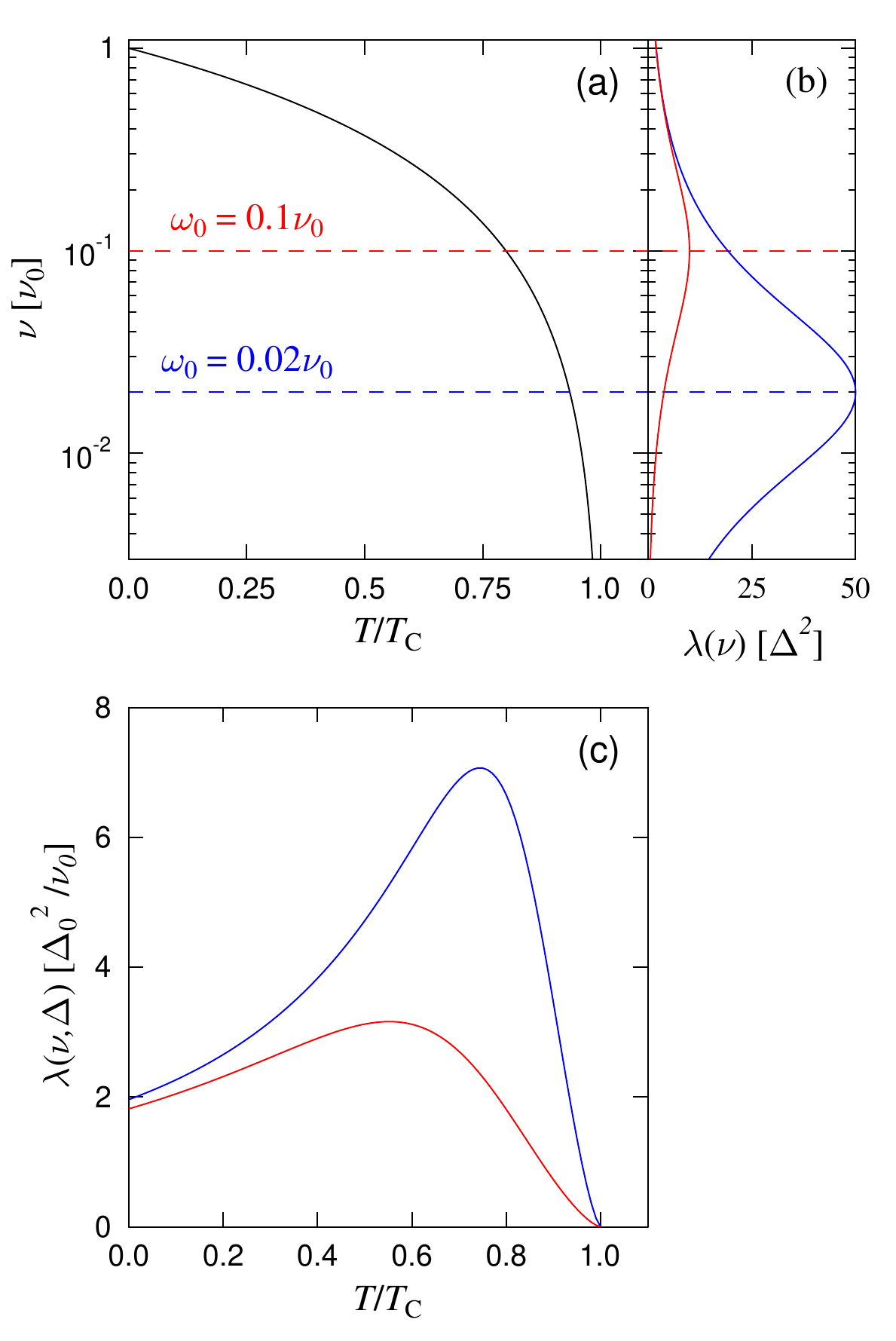}
\caption{Schematic showing the expected behavior of the $\mu$SR relaxation rate $\lambda$ from a fluctuation with a frequency that decreases approaching $T_C$. (a) Fluctuation rate as a function of temperature, modelled using a 3D Heisenberg power law $\nu = \nu_0 \left(1-\frac{T}{T_\text{c}}\right)^{\zeta z}$, where $\nu_0$ is the zero-temperature frequency, $\zeta=0.7048$ is the correlation-length critical exponent, and $z=2.035$ is the dynamical critical exponent \cite{Hicken2020b}. (b) Relaxation rate ($x$-axis) as a function of frequency ($y$-axis) following the Redfield formula, for the case of constant $\Delta$, for two different values of the applied field ($\omega_0 = \gamma_{\mu} B_{\mathrm{L}}$). (c)  Relaxation rate as a function of temperature, also now including a power-law decrease in $\Delta$ with increasing temperature. \cite{Hicken2020}}
\label{fig:schematic}
\end{figure}

The data measured at 0.01~T in Fig.~\ref{fig:MnNiGa}(b) show a  peak in the $\mu$SR relaxation rate $\lambda$ at 340~K, along with a step-like drop in the relaxing asymmetry typical of a transition to long-range order. These features occur at the same temperature as the minimum in d$M$/d$T$ that marks $T_{\mathrm{C}}$ in the magnetometry data.
In the fast-fluctuation limit, the $\mu$SR relaxation rate $\lambda$ is often described by the Redfield formula \cite{Yaounc2010},
\begin{equation}
\lambda = \frac{2\Delta^2\nu}{{\omega_0}^2 + \nu^2},
\end{equation}
where $\Delta = \gamma_{\mu}^{2}\sqrt{\langle \delta B^{2}\rangle} $ is determined by the amplitude of the fluctuations and $\nu$ is a fluctuation rate.
If the relaxation is determined by fluctuations with a characteristic fluctuation rate $\nu$ that decreases as the transition temperature is approached from below, then
the $\mu$SR relaxation rate $\lambda$ would be expected to show a peak slightly below $T_C$ which broadens and shifts to lower temperature as the applied magnetic field is increased, as illustrated by Fig.~\ref{fig:schematic} \cite{Hicken2020}. Our data follows this general trend of behaviour, with the $\mu$SR data at larger magnetic fields showing a high temperature peak in the relaxation rate that is broader, weaker, and shifted to lower temperatures compared to the 0.01~T data.

However, a potentially puzzling feature occurs in the imaginary part of the AC magnetic susceptibility [$\chi ''$, green line, lower panel of Fig.~\ref{fig:MnNiGa}(b)], which shows a sharp peak
15~K lower in temperature than the peak in the $\mu$SR relaxation for the 0.01~T dataset. AC susceptibility probes a substantially lower frequency range than $\mu$SR, and therefore we might expect the influence of fluctuations with a single characteristic fluctuation rate that decreases with increasing $T$  to appear at higher temperatures in AC than in the $\mu$SR. The observation of a peak in  $\chi ''$ at lower temperature therefore
suggests

that there are spin fluctuations leading to 
additional dynamics at frequencies that are too low to be seen with $\mu$SR.

Furthermore, there is no obvious signature of the high temperature transition in $\chi''$ at any of the fields above 0.01~T.
It is possible that this simply arises from a broadening of the transition with field, which would reduce the measured peak height of the imaginary signal, potentially bringing it below the noise floor. However, since Fig.~\ref{fig:MnNiGa}(a) indicates that at these higher fields the system is in a different magnetic state, it may instead be that the low-frequency spectral weight that gives rise to the peak in $\chi''$ at 0.01~T does not persist into these higher magnetic field states.

At lower temperature in the $\mu$SR data, we see a further drop in the relaxing asymmetry at 200~K, which coincides with a minimum in d$M$/d$T$. This is consistent with the temperature of the spin-reorientation transition to a non-colinear canted ferromagnetic state that has been suggested from neutron scattering measurements \cite{Xu2019}.
The additional drop in the relaxing asymmetry would, for example, occur if a larger fraction of the static magnetic moments in the material are canted away from the applied field direction below 200~K, which would increase the fast-relaxing component of the $\mu$SR signal,  leading to the additional missing asymmetry we observe.
This transition does not coincide with a peak in the $\mu$SR relaxation rate, nor a peak in the imaginary part of the magnetic susceptibility. Instead, there is a broad increase in both of these quantities below the 200~K transition,
suggesting that  the low-temperature dynamics occur over a wide range of frequencies (from 10s of Hz to MHz) and persist across a broad temperature range. As alluded to by Bogdanov \textit{et al.} \cite{Bogdanov2002},  this situation can arise from a near-degeneracy in canting angles around the field direction, resulting in a small energy barrier between magnetic domains that would therefore undergo significant dynamics. 

We  see limited qualitative difference in the $\mu$SR behavior crossing from 0.2~T to 0.3~T, where the previous reports suggest a transition from the helical or stripe state to the biskyrmion state. We also do not see evidence for a transition between these states as a function of temperature, as might be expected to occur between 250 and 350~K for the 0.2 T dataset.
This suggests that the proposed transition from the helical to biskyrmion states does not produce a sizeable enough change to the local magnetic fields or their dynamics to give a resolvable response in the $\mu$SR data, which  contrasts with previous $\mu$SR measurements of Bloch and N\'{e}el type skyrmions \cite{Stefancic2018,Franke2018,Hicken2020}, where the presence of skyrmions is detectable. Therefore, our $\mu$SR data suggest that either the biskyrmion state does not host characteristic dynamics similar to those of conventional skyrmions, or  this biskyrmion state does not persist in bulk samples, as might be expected if the interpretation of biskyrmions as type-II bubbles stabilized by demagnetization and confinement effects is correct \cite{Loudon2019, Turnbull2020}.

\subsection{Mn$_{1.4}$Pt$_{0.9}$Pd$_{0.1}$Sn}
\subsubsection{Magnetometry}
\begin{figure}[h!]
\includegraphics[width=0.75\columnwidth]{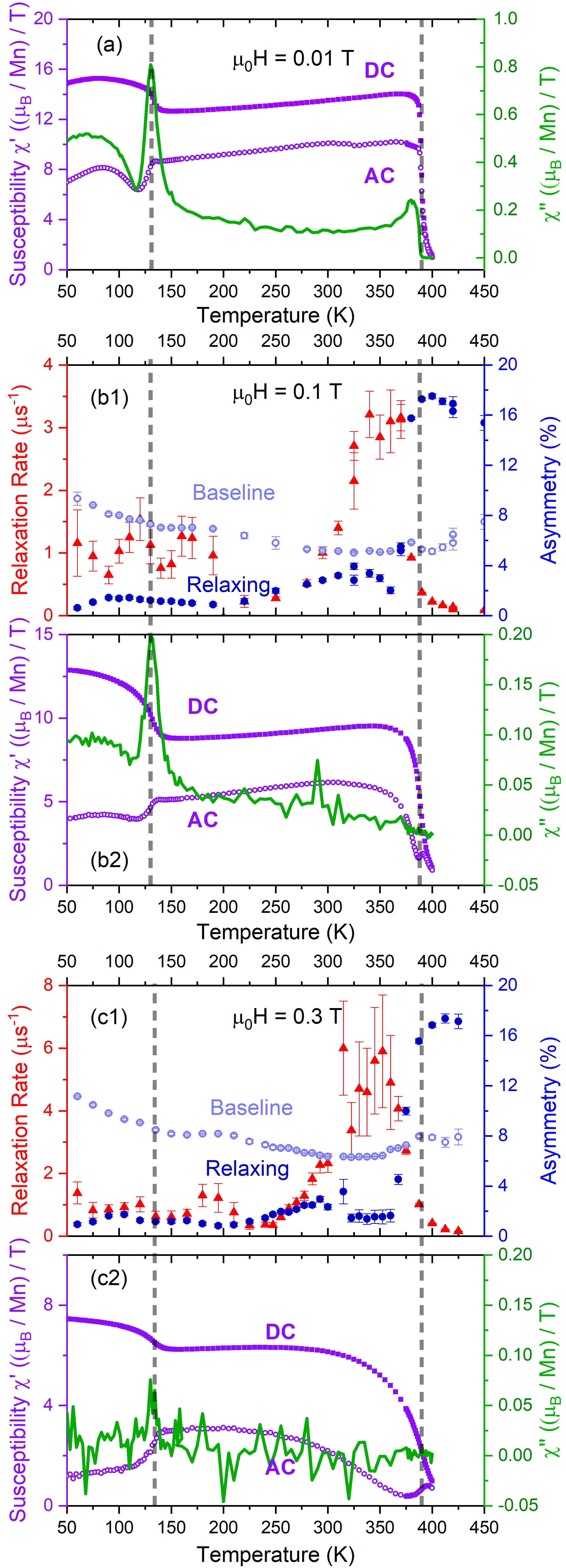}
\caption{(a) Magnetometry data for \antisky~measured at 0.01~T. $\mu$SR and magnetometry data measured at (b) 0.1~T and (c) 0.3~T after zero-field cooling. Panel (1) shows  $\mu$SR relaxation rate  (red triangles, left axis),  baseline asymmetry (open light blue circles, right axis), relaxing asymmetry  (blue circles, right axis). Panel (2) shows  DC magnetization plotted as $M/H$  (purple squares, left axis),  real part of the AC susceptibility ($\chi'$)  (purple open circles, left axis), and  imaginary part of the AC susceptibility ($\chi''$)  (green solid line, right axis). Gray dashed lines indicate transition temperatures determined from minima in d$M$/d$T$ at each magnetic field.}
\label{fig:Antiskyrmion}
\end{figure}

We now turn to discussion of the data measured on the reported antiskyrmion host,
\antisky. The DC magnetization  in Fig.~\ref{fig:Antiskyrmion} shows two features at each field: a sharp rise with decreasing temperature occurring around 390~K, and a second rise starting at 145~K. This indicates the presence of two separate magnetic transitions in the material, at $T_{\textrm{C}} \approx 380$~K and $T_{\textrm{SR}} \approx 140$~K, consistent with previous reports \cite{Nayak2017}. The rise in magnetization at the lower-temperature transition, in contrast to MnNiGa, shows that the spin-reorientation transition involves an increase in the ferromagnetic component of the magnetism at lower temperature. The AC susceptibility also shows two features coinciding with the two transitions, although these are slightly different at the different magnetic fields. At 0.1 and 0.3~T, $T_{\textrm{C}}$ is identified by a small peak in $\chi'$ with no feature in $\chi''$, while $T_{\textrm{SR}}$ shows up as a step-like drop in $\chi'$ and a peak in $\chi''$. 
At 0.01~T, both transitions show a peak in $\chi''$. This indicates that the lower temperature transition shows dynamics with frequencies close to the AC drive field of 111~Hz at all fields. At temperature below $T_{\textrm{SR}}$, $\chi''$ continues to increase down to low temperature in the 0.1 and 0.01~T data, which suggests that there are spin dynamics
at low $T$
that are at least partially suppressed by the application of a 0.3~T magnetic field. 

\subsubsection{$\mu$SR}

\begin{figure}
\includegraphics[width=\columnwidth]{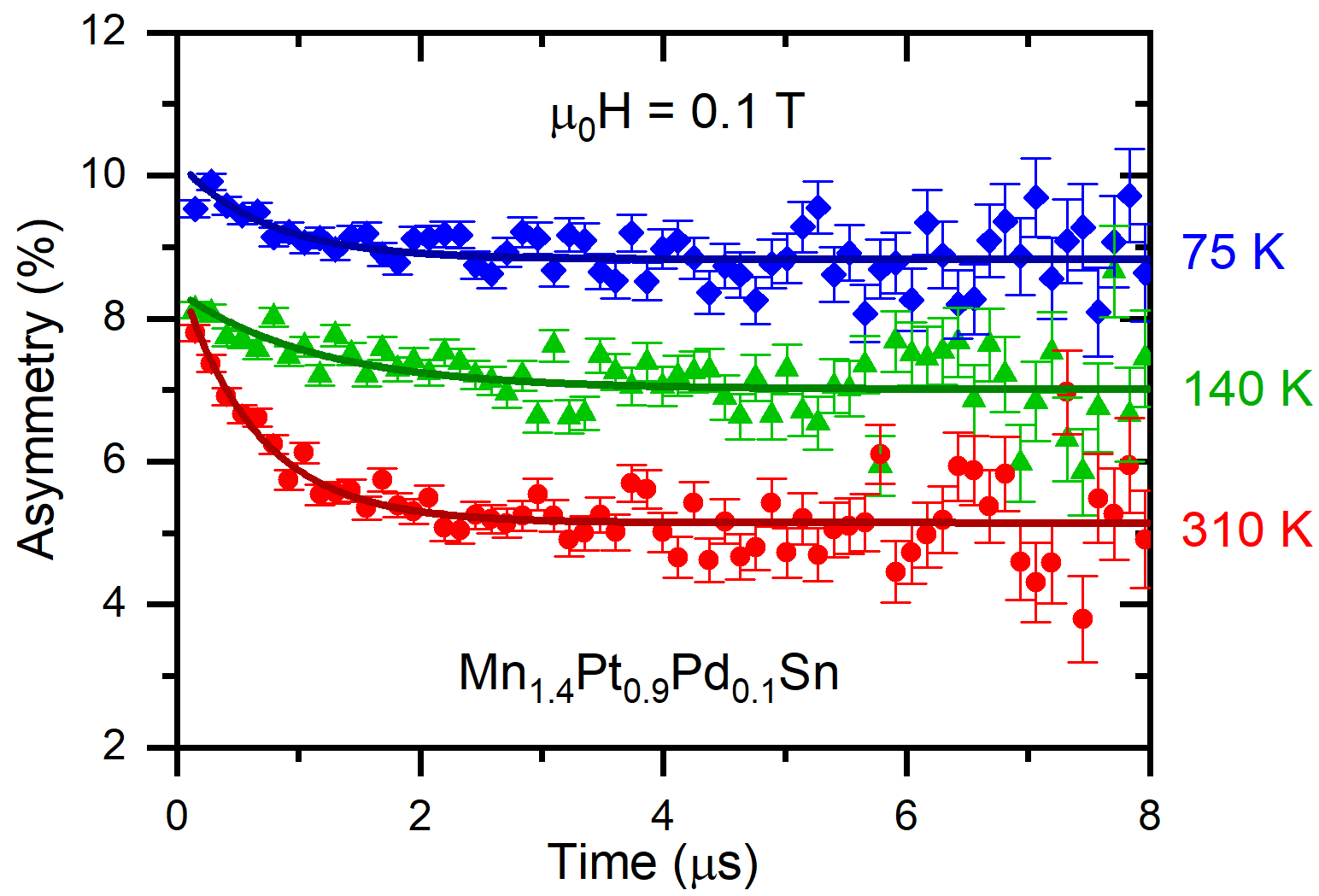}
\caption{Representative $\mu$SR data measured in longitudinal fields of 0.1~T for \antisky~at 75~K, 140~K, and 310~K. Solid points show measured data; lines show fits to Eq.~\ref{eq:1}.}
\label{fig:data_Antisky}
\vspace{-1cm}
\end{figure}

Figure~\ref{fig:data_Antisky} shows representative $\mu$SR data measured in a longitudinal magnetic field of 0.1~T at several temperatures. These data again show exponential relaxation that can be modelled by Eq.~\ref{eq:1}, with the notable difference compared to the MnNiGa data that the non-relaxing baseline ($A_\textrm{b}$) in these data varies as a function of temperature. Baseline asymmetry varying with temperature is unusual in $\mu$SR, as this term is normally temperature-independent and arises from muons hitting the silver sample holder (which show no relaxation in a longitudinal field greater than about 5~mT). We therefore fit the $\mu$SR data to Eq.~\ref{eq:1}, allowing the baseline, relaxation rate, and relaxing asymmetry to freely vary with temperature. The results of this fitting procedure are shown in Fig.~\ref{fig:Antiskyrmion} for fields of 0.1~T and 0.3~T, along with magnetometry data at the same magnetic fields for comparison, and magnetometry data measured at 0.01~T as a reference.

We see evidence for the two magnetic transitions in the $\mu$SR data shown in Fig.~\ref{fig:Antiskyrmion}(b1) and (c1). The higher-temperature transition is characterized by a large peak in the relaxation rate at around 360~K, along with a sharp drop in the relaxing asymmetry. This peak is centered $\approx 30$~K below $T_{\textrm{C}}$, and is not accompanied by a peak in $\chi''$. This broad peak suggests that there are dynamics in the system below $T_{\textrm{C}}$ with a fluctuation rate that decreases slowly in frequency as the temperature is increased towards $T_{\textrm{C}}$, hence putting its characteristic frequency within the $\mu$SR time window over a broad temperature range, similar to the picture shown in  Fig.~\ref{fig:schematic}.

It has been reported from magnetic entropy measurements that bulk \antisky~hosts antiskyrmions up to 350~K or higher at 0.1~T, while it does not host them until below 300~K at 0.3~T \cite{Jamaluddin2019}. It is plausible that near the higher-temperature transition we are probing dynamics arising from characteristic excitations of the antiskyrmion state at this lower field. However, this would require that the antiskyrmion state is characterized by a lower relaxation rate than the neighbouring magnetic states, in contrast with published data on Bloch and Neel skyrmions, where an increase in this quantity is seen \cite{Stefancic2018,Franke2018,Hicken2020,Hicken2020b}.

At lower temperaures, there are further features in the $\mu$SR data: two small peaks in the relaxation rate, around 175~K and 100~K for both fields,
which roughly correspond with kinks in the baseline asymmetry, and a rise in the relaxation rate at the lowest measured temperatures which is correlated with a further drop in the relaxing asymmetry. The behavior of the relaxing asymmetry in particular is quite distinct to that in MnNiGa, showing that these transitions behave differently, despite both being spin reorientations, with the complicated temperature dependence in \antisky~around the low-temperature transition suggesting a multi-step evolution of the dynamics of the magnetism at these temperatures.

Furthermore, the continued rise in the imaginary part of the susceptibility below $T_{\textrm{SR}}$, and enhancement in the $\mu$SR relaxation rate at the lowest temperatures, suggests that there are  low-temperature spin dynamics across a range of frequencies in this low temperature phase, similar to what was seen in MnNiGa.
Recent reports have suggested that there is a complicated evolution between different geometries of bubbles or antiskyrmions as a function of temperature and magnetic field in thin plates of \antisky~\cite{Jena2020-2, Peng2020}, with circular bubbles, elliptical bubbles, triangular bubbles, and square `antiskyrmions' all appearing at different points of the phase diagram. Transitions between these different states as a function of temperature may explain the multi-step evolution of the dynamics indicated by our $\mu$SR data.

\begin{figure}[t]
\includegraphics[width=0.75\columnwidth]{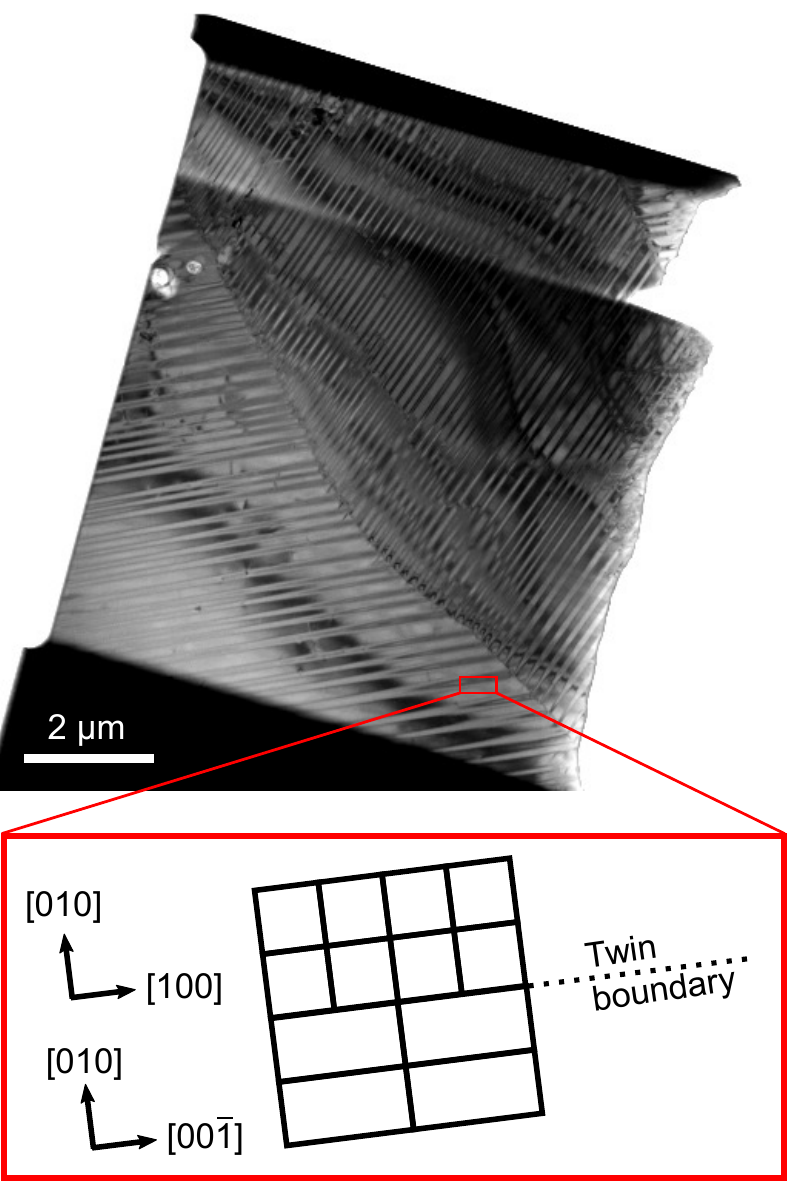}
\caption{Bright field TEM Image of \antisky~showing the presence of twinning domains in the crystal structure.}
\label{fig:TEM}
\vspace{-0.2cm}
\end{figure}

The other notable feature at low temperatures is the continued rise in the baseline asymmetry with decreasing temperature, with a plateau approximately between the two low temperature peaks in the relaxation rate. This feature likely arises from a combination of two factors. First, if we have some preferred orientation in our polycrystalline sample (as is likely), and the spin reorientation transition causes the internal magnetic field to more frequently point along the initial muon polarization direction due to this preferred orientation, there will be a reduction in the static contribution to the muon signal (which is seen as missing initial asymmetry in our data). This would result in an increasing initial asymmetry, as is seen between the 140~K and 75~K data sets in Fig. \ref{fig:data_Antisky}. However, in this case, regions of the sample where the internal field lies parallel to the muon spin would still show dynamic relaxation caused by fluctuations of the moments. Our data shows both a non-relaxing contribution to the muon asymmetry, and a contribution with relaxation arising from dynamics. This suggests that there are spatially separated regions of the sample that show different dynamics: regions where the dynamics are either too fast or too slow to be probed by $\mu$SR (giving a non-relaxing signal), and regions where dynamics fall into the $\mu$SR frequency window (giving a relaxing signal). Since the total change in the baseline asymmetry with temperature is approximately 5\%, and the maximum relaxing asymmetry is 20\%, we can infer that these regions with dynamics outside the $\mu$SR frequency window occupy approximately a quarter of the sample volume. It is possible that these different regions arise from effects near structural crystal domain boundaries. Notably, this material is known to be prone to twinning defects \cite{Nayak2017}, and transmission electron microscopy (TEM) images of thin lamella taken from of our sample(see Fig. \ref{fig:TEM}) demonstrate that we have closely spaced ($<50$~nm) structural twinning defects throughout our sample. Magnetism varying slightly surrounding these defects (as previously reported \cite{Nayak2017}) would result in a substantial portion of the sample experiencing different dynamics, explaining our observations.

\section{Conclusion}

In conclusion, we have presented $\mu$SR and magnetometry of bulk samples of two materials (MnNiGa and \antisky) thought to host skyrmionic states in thin lamellae form, which show bulk spin reorientation transitions with temperature. In both samples, we find evidence for two magnetic transitions as a function of temperature. Our data suggests that the higher-temperature transition in both samples is characterized by dynamics whose 
fluctuation rate
 decreases in frequency with increasing temperature.
In both materials, we also find evidence for spin dynamics persisting down to low temperatures below the spin-reorientation temperatures. However, the spin re-orientation transitions themselves show rather different signatures in the two systems, with \antisky~showing an increase in the ferromagnetic moment and the peak in $\chi ''$
but also suggesting some spatial variation of dynamics across the sample from $\mu$SR, while the MnNiGa data suggests a more continuous change in dynamics across this transition.
Finally, we note that the $\mu$SR measurements do not show unambiguous evidence for dynamics reflect ting the proposed biskyrmion state in MnNiGa, which would suggest that either this state does not occur in bulk samples, or that its dynamics  are different from those seen in other skyrmionic systems, and are not distinguishable from those of the low-temperature helical or stripe phase.
However, this work does further advances our understanding of the behavior of the bulk materials, which underpins the observation of topological magnetism in thin lamellae.

Data presented in this paper are available at \cite{doi}.

\begin{acknowledgements}
Experiments at the ISIS Pulsed Neutron and Muon Source were supported by a beamtime allocation from the Science and Technology Facilities Council. This work was supported by the UK Skyrmion Project EPSRC Programme Grant (EP/N032128/1). M.~N.~Wilson acknowledges the support of the Natural Sciences and Engineering Research Council of Canada (NSERC). M. Gomil\v{s}ek acknowledges the support of the Slovenian Research Agency under Project No. Z1-1852. 
\end{acknowledgements}


\begin{thebibliography}{2}
\bibitem{Nagaosa2013} N. Nagaosa, and Y. Tokura, \textit{Topological properties and dynamics of magnetic skyrmions}, Nature Nanotechology \textbf{8}, 899 (2013).

  \bibitem{Lancaster2019} T. Lancaster, \textit{Skyrmions in magnetic materials}, Contemporary Physics, \textbf{60}, 246 (2019).

\bibitem{Yu2014} X.Z. Yu, Y. Tokunaga, Y. Kaneko, W. Z. Zhang, K. Kimoto, Y. Matsui, Y. Taguchi, and Y. Tokura, \textit{Biskyrmion states and their current-driven motion in a layered manganite}, Nature Communications \textbf{5}, 3198 (2014).


\bibitem{Nayak2017} A. K. Nayak, V. Kumar, T. Ma, P. Werner, E. Pippel, R. Sahoo, F. Damay, U. K. R\"{o}\ss ler, C. Felser, and S.S.P. Parkin, \textit{Magnetic antiskyrmions above room temperature in tetragonal Heusler materials}, Nature \textbf{548}, 561 (2017).


\bibitem{Peng2017} L. Peng, Y. Zhang, W. Wang, M. He, L. Li, B. Ding, J. Li, Y. Sun, X.-G. Zhang, J. Cai, S. Wang, G. Wu, and B. Shen, \textit{Real-Space Observation of Nonvolatile Zero-Field Biskyrmion Lattice Generation in MnNiGa Magnet}, Nano Letters \textbf{17}, 7075 (2017).

\bibitem{Wang2016} W. Wang, Y. Zhang, G. Xu, L. Peng, B. Ding, Y. Wang, Z. Hou, X. Zhang, X. Li, S. Wang, J. Cai, F. Wang, J. Li, F. Hu, G. Wu, B. Shen, and X.-X. Zhang, \textit{A Centrosymmetric Hexagonal Magnet with Superstable Biskyrmion Magnetic Nanodomains in a Wide Temperature Range of 100–340 K}, Advanced Materials, \textbf{28}, 6887 (2016).

  
\bibitem{Jena2020} J. Jena, R. Stinshoff, R. Saha, A.K. Srivastava, T. Ma, H. Deniz, P. Werner, C. Felser, and S. P. Parkin, \textit{Observation of Magnetic Antiskyrmions in the Low Magnetization Ferrimagnet Mn$_2$Rh$_{0.95}$Ir$_{0.05}$Sn}, Nano Lett. \textbf{20}, 59 (2020).
\bibitem{Kumar2020} V. Kumar, N. Kumar, M. Reehuis, J. Gayles, A.S. Sukhanov, A. Hoser, F. Damay, C. Shekhar, P. Adler, and C. Felser, \textit{Detection of antiskyrmions by topological Hall effect in Heusler compounds}, Phys. Rev. B \textbf{101}, 014424 (2020).

\bibitem{Swekis2019} P. Swekis, A. Markou, D. Kriegner, J. Gayles, R. Schlitz, W. Schnelle, S. T. B. Goennenwein, and C. Felser, \textit{Topological Hall Effect in thin films of Mn$_{1.5}$PtSn}, Phys. Rev. Matt. \textbf{3}, 013001(R) (2019).
\bibitem{Rana2016} K.G. Rana, O. Meshcheriakova, J. \"{u}bler, B. Ernst, J. Karel, R. Hillebrand, E. Pippel, P. Wener, A. K. Nayak, and C. Felser, \textit{Observation of topological Hall effect in Mn$_{2}$RhSn films}, New Journal of Physics \textbf{18}, 085007 (2016).
\bibitem{Liu2018} Z.H. Liu, Burigu A., Y.J. Zhang, H.M. Jafri, X.Q. Ma, E.K. Lie, W.H. Wang, and G.H. Wu, \textit{Giant topological Hall effect in tetragonal Heusler alloy Mn$_{2}$PtSn}, Scripta Materialia \textbf{143}, 122 (2018).
\bibitem{Li2018} Y. Li, B. Ding, X. Wang, H. Zhang, W. Wang, and Z. Liu, \textit{Large topological hall effect observed in tetragonal Mn$_{2}$PtSn Heusler thin film}, Appl. Phys. Lett. \textbf{113}, 062406 (2018).
\bibitem{Loudon2019} J.C. Loudon, A.C. Twitchett-Harrison, S. Cort\'{e}s-Ortu\~{n}o, M.T. Birch, L.A. Turnbull, A. {\v S}tefan{\v c}i{\v c}, F.Y. Ogrin, E.O. Burgos-Parra, N. Bukin, A. Laurenson, H. Popescu, M. Beg, O. Hovorka, H. Fangohr, P. A. Midgley, G. Balakrishnan, and P. D. Hatton, \textit{Do Images of Biskyrmions Show Type\- II Bubbles?}, Advanced Materials, \textbf{31}, 1806598 (2019).

\bibitem{Capic2019} D. Capic, D.A. Garanin, and E. M. Chudnovsky. \textit{Biskyrmion lattices in centrosymmetric magnetic films}, Physical Review Research, \textbf{1}, 033011 (2019).


\bibitem{Vir2019} P. Vir, N. Kumar, H. Borrmann, B. Jamijansuren, G. Kreiner, C. Shekhar, and C. Felser, \textit{Tetragonal Superstructure of the Antiskyrmion Hosting Heusler
Compound Mn$_{1.4}$PtSn}, Chem. Mater. \textbf{31}, 5876 (2019).
\bibitem{Meshcheriakova2014} O. Meshcheriakova, S. Chadov, A.K. Nayak, U. K. R\"{o}\ss ler, J. K\"{u}bler, G. Andr\'{e}, A. A. Tsirlin, J. Kiss, S. Hausdorf, A. Kalache, W. Schnelle, M. Nicklas, and C. Felser, \textit{Large Noncollinearity and Spin Reorientation in the Novel Mn$_2$RhSn Heusler Magnet}, Phys. Rev. Lett. \textbf{113}, 087203 (2014).

\bibitem{Shiraishi1999} H. Shiraishi, H. Niida, Y. Iguchi, S. Mitsudo, M. Motokawa, K. Ohayama, H. Miki, H. Onodera, T. Hori, and K. Kanematsu, \textit{Structural and magnetic properties of Ni$_2$In type (Mn$_{1-x}$Ni$_x$)$_{65}$Ga$_{35}$ compounds}, Journal of Magnetism and Magnetic Materials \textbf{196-197}, 660 (1999).
\bibitem{Xu2019} G. Xu, Y. You, J. Tang, H. Zhang, H. Li, X. Miao, Y. Gong, Z. Hou, Z. Cheng, J. Wang, A. J. Studer, F. Xu, and W. Wang, \textit{Simultaneous tuning of magnetocrystalline anisotropy and spin reorientation transition via Cu substitution in Mn-Ni-Ga magnets for nanoscale biskyrmion formation}, Phys. Rev. B \textbf{100}, 054416 (2019).

\bibitem{Qian2016} F. Qian, H. Wilhelm, A. Aqeel, T.T.M. Palstra, A.J.E. Lefering, E. H. Br\"{u}ck, and C. Pappas, Phys. Rev. B \textbf{94}, 064418 (2016).

\bibitem{Yaounc2010} A. Yaouanc, and P. Dalmas de R\'{e}otier, \textit{Muon Spin Rotation, Relaxation, and Resonance: Applications to Condensed Matter (International Series of Monographs on Physics)}, Oxford Science Publications, (2010).


\bibitem{Bogdanov2002} A.N. Bogdanov, U.K. R\"{o}\ss ler, M. Wolf, and K.-H. M\"{u}ller, \textit{Magnetic Structures and reorientation transitions in noncentrosymmetric uniaxial antiferromagnets}, Phys. Rev. B \textbf{66}, 214410 (2002).

\bibitem{Stefancic2018}
A.\v{S}tefan\v{c}i\v{c}, S. H. Moody, T.J. Hicken, M. T. Birch, G. Balakrishnan, S. A. Barnett, M. Crisanti, J. S. O. Evans, S. J. R. Holt, K. J. A. Franke, P. D. Hatton, B. M. Huddart, M. R. Lees, F. L. Pratt, C. C. Tang, M. N. Wilson, F. Xiao, and T. Lancaster, \textit{Origin of skyrmion lattice phase splitting in Zn-substituted Cu$_2$OSeO$_3$}, Physical Review Materials \textbf{2}, 111402(R) (2018).
\bibitem{Franke2018}
K.J. A. Franke, B. M. Huddart, T. J. Hicken, F. Xiao, S. J. Blundell, F. L. Pratt, M. Crisanti, J. A. T. Barker, S. J. Clark, S. \v{S}tefan\v{c}i\v{c}, M. C. Hatnean, G. Balakrishnan, and T. Lancaster, \textit{Magnetic phases of skyrmion-hosting GaV$_4$S$_{8-y}$Se$_y$ probed with muon spectroscopy}, Physical Review B \textbf{98}, 054428 (2018).
\bibitem{Hicken2020}
T. J. Hicken, S.J.R. Holt, K.J.A. Franke, Z. Hawkhead, S. \v{S}tefan\v{c}i\v{c}, M. N. Wilson, M. Gomil\v{s}ek, B.M. Huddart, S. J. Clark, M. R. Lees, F. L. Pratt, S. J. Blundell, G. Balakrishnan, and T. Lancaster, \textit{Magnetism and N\'{e}el skyrmion dynamics in GaV$_{4}$S$_{8-y}$Se$_y$}, Physical Review Research \textbf{2}, 032001(R) 2020.

\bibitem{Hicken2020b}
T. J. Hicken, M.N. Wilson, K.J.A. Franke, B.M. Huddart, Z. Hawkhead, M. Gomil\v{s}ek, S.J. Clark, F.L. Pratt, A. S. \v{S}tefan\v{c}i\v{c}, A. E. Hall, M. Ciomaga Hatnean, G. Balakrishnan, T. Lancaster, \textit{Megahertz dynamics in skymion systems probed with muon-spin relaxation}, Phys. Rev. B \textbf{103}, 024428 (2021).

\bibitem{Turnbull2020}
L. A. Turnbull, M. T. Birch, A. Laurenson, N. Bukin, E. O. Burgos-Parra, H. Popesco, M. N. Wilson, A. Stefan\v{c}i\v{c}, G. Balakrishnan, F. Y. Ogrin, and P. D. Hatton, \textit{Tilted X-Ray Holography of Magnetic Bubbles in MnNiGa Lamellae}, ACS Nano 1936-0851 (2020).


\bibitem{Jamaluddin2019} S. Jamaluddin, S.K. Manna, B. Giri, P.V.P. Madduri, S.S.P. Parkin, and A.K. Nayak, \textit{Robust Antiskyrmion Phase in Bulk Tetragonal  Mn-Pt(Pd)-Sn Heusler System Probed by Magnetic Entropy Change and AC-Susceptibility Measurements}, Advanced Functional Materials \textbf{29}, 1901776 (2019).

\bibitem{Jena2020-2} J. Jena, B. G\"{o}bel, T. Ma, V. Kumar, R. Saha, I. Mertig, C. Felser, and S.S.P. Parkin, \textit{Elliptical Bloch Skyrmion chiral twins in an antiskyrmion system}, Nature Communications \textbf{11}, 1115 (2020).

\bibitem{Peng2020} L. Peng, R. Takagi, W. Koshibae, K. Shibata, K. Nakajima, T.-h. Arima, N. Nagaosa, S. Seki, X. Yu, and Y. Tokura, \textit{ Controlled transformation of skyrmions and antiskyrmions in a non-centrosymmetric magnet}, Nature Nanotechnology \textbf{15}, 181 (2020).

\bibitem{doi} doi:10.15128/r13j333226f 


\end{thebibliography}
\end{document}